\documentclass[final,3p,times,twocolumn]{elsarticle} 
\usepackage{amssymb,color}

\journal{Journal of Physics and Chemistry of Solids}

\begin{document}

\begin{frontmatter}

\title{Effect of external pressure on the magnetic properties\\ of $R$CoAsO ($R$ = La, Pr, Sm): a $\mu$SR study}

\author[ifw]{G.~Prando}
\author[upv]{S.~Sanna}
\author[psi]{R.~Khasanov}
\author[npl]{A.~Pal}
\author[ifw]{E.~M.~Br\"uning}
\author[upr]{M.~Mazzani}
\author[npl]{V.~P.~S.~Awana}
\author[ifw,ift]{B.~B\"uchner}
\author[upr]{R.~De~Renzi\corref{cor1}}

\cortext[cor1]{Corresponding author: roberto.derenzi@unipr.it}
\address[ifw]{Leibniz-Institut f\"ur Festk\"orper- und Werkstoffforschung (IFW) Dresden, D-01171 Dresden, Germany}
\address[upv]{Dipartimento di Fisica and Unit\`a CNISM di Pavia, Universit\`a di Pavia, I-27100 Pavia, Italy}
\address[psi]{Laboratory for Muon Spin Spectroscopy, Paul Scherrer Institut, CH-5232 Villigen PSI, Switzerland}
\address[npl]{National Physical Laboratory (CSIR), New Delhi 110012, India}
\address[upr]{Dipartimento di Fisica and Unit\`a CNISM di Parma, Universit\`a di Parma, I-43124 Parma, Italy}
\address[ift]{Institut f\"ur Festk\"orperphysik, Technische Universit\"at Dresden, D-01062 Dresden, Germany}

\begin{abstract}
We report on a detailed investigation of the itinerant ferromagnets LaCoAsO, PrCoAsO and SmCoAsO performed by means of muon spin spectroscopy upon the application of external hydrostatic pressures $p$ up to $2.4$ GPa. These materials are shown to be magnetically hard in view of the weak dependence of both critical temperatures $T_{C}$ and internal fields at the muon site on $p$. In the cases $R$ = La and Sm, the behaviour of the internal field is substantially unaltered up to $p = 2.4$ GPa. A much richer phenomenology is detected in PrCoAsO instead, possibly associated with a strong $p$ dependence of the statistical population of the two different crystallographic sites for the muon. Surprisingly, results are notably different from what is observed in the case of the isostructural compounds $R$CoPO, where the full As/P substitution is already inducing a strong chemical pressure within the lattice but $p$ is still very effective in further affecting the magnetic properties.
\end{abstract}


\date{\today}

\end{frontmatter}

\section{Introduction}

The mutual interaction among localized $f$ and itinerant $d$ electrons in the quaternary compounds $RBC$O (or, shortly, $1111$) has attracted interest for these materials \cite{Kre07,Kre08} well before the discovery of high-$T_{c}$ superconductivity in F-doped $R$FeAsO \cite{Kam08,Joh10,Ste11}. Here, the crystallographic structure is composed of alternating layers of $R$O (where, typically, $R$ is a rare-earth ion) and $BC$ (with $B$ a transition-metal ion and $C$ a pnictogen element) \cite{Kam08,Gar08,Mar09}. O ($B$) ions are surrounded by $R$ ($C$) ions in a tetrahedral environment within each layer \cite{Kam08,Gar08,Mar09}. As a result of the coexistence of $f$ and $d$ electrons, peculiar magnetic features emerge in $1111$ materials. As a well-known example, a spin density wave (SDW) phase coexists with an antiferromagnetically (AFM) ordered phase of $R$ magnetic moments for $T \lesssim 5 - 10$ K in $R$FeAsO \cite{Mae09,Shi11,DeR12,Mae12,Pra13b}. Besides the charge doping, achieved by means of O$_{1-x}$F$_{x}$ or Fe$_{1-x}$Co$_{x}$ chemical substitutions, external pressure is also known to lead to a superconducting ground state in $R$FeAsO, even for undoped compounds (see \cite{Chu09} and references therein). In this respect, it should be stressed that the chemical substitution of $R$ ions also shrinks the lattice and sizeably affects crystallographic properties like the tetrahedral angle and the pnictogen height, two quantities which are strictly correlated with the value of the superconducting transition temperature $T_{c}$ \cite{Miy13}. Accordingly, only small $R$ ions like Sm allow to obtain the maximum $T_{c}$ value of $\sim 55$ K -- incidentally, the highest value known up to now for pnictide materials \cite{Ren08,Pra10,Pra11}.

A more complex magnetic behaviour is detected in $R$CoAsO materials \cite{Yan08,Oht09a,Oht09b,Oht10a,Oht10b,Sar10,Awa10b,Pal11a,Oht11,Sug11,Pal11b,Maj13}. Here, an itinerant ferromagnetic (FM) phase is achieved below $T_{C} \simeq 60 - 80$ K, the precise value of $T_{C}$ being again dependent on $R$ \cite{Yan08,Oht09b,Oht10a,Oht11,Sug11,Maj13,Pra13a}. At lower temperatures, the occurrence of progressive FM to AFM transitions of the Co sublattice induced by magnetic $R$ ions is observed \cite{Oht09b,Oht10a,Oht11,Sug11}. As already stressed above, $R$ ions influence the magnetic structure of $R$CoAsO materials not only through their weak exchange couplings, both with Co and within their own sublattice, but also through the distortion that the ion size mismatch induces. This second effect is often described as chemical pressure, to underline its analogy with the application of external pressure. The trend described in Ref.~\cite{Sug11} is that by decreasing the ionic radius from La to Gd, i.~e., by shrinking the lattice in analogy to what external pressure may do, there is a tendency to replace FM with AFM order at low temperatures, with an increasing $T_{N}$. In particular, muon spin spectroscopy ($\mu$SR) data suggest that La, Ce and Pr are FM throughout their ordered phase, whereas Nd, Sm and Gd undergo a FM to AFM transition at an intermediate critical temperature $T_{N}$. In the case of $R$ = Pr, unspecified intermediate transitions within the FM phase are suggested as a viable explanation for discontinuous features in the $\mu$SR data \cite{Sug11}. However, a recent work indicates that the Pr moment does not participate to the magnetic order down to $5$ K \cite{Tiw14}.

In a previous $\mu$SR experiment on a related series of isostructural rare-earth cobaltates $R$CoPO ($R$ = La, Pr), a strong sensitivity to pressure was detected. In particular, we determined that in those samples the effect of chemical and external pressures is very similar \cite{Pra13a}. A strikingly similar modification of the local field at the muon is observed both when changing $R$ from La to Pr and when applying pressure on each individual sample, while magnetization measurements demonstrate that the local moment on Co is not changing significantly in either cases \cite{Pra13a}. DFT calculations indicate that the muon interstitial site does not change upon applying pressure and that the effect can be explained by a Fermi contact hyperfine field originating from a band approaching and finally crossing the Fermi surface, both when external pressure is increased and when the smaller Pr ions are chosen \cite{Pra13a}.

To further elucidate the influence of pressure (both external hydrostatic and chemical) on the magnetic order of Co in the $1111$ structure as well as on the muon probe itself, we report here the extension of our $\mu$SR investigation to three samples of $R$CoAsO ($R$ = La, Pr, Sm). As mentioned above, the ``magnetic softness'' of $R$CoPO was reported to be sizeable. Since As is introducing less chemical pressure than P, an even stronger sensitivity to $p$ would be naively expected for $R$CoAsO. Surprisingly, as main output of our measurements, $R$CoAsO are ``magnetically hard'' materials since no remarkable effect is induced by pressure up to $p = 2.4$ GPa. Both LaCoAsO and SmCoAsO are virtually insensitive to $p$, besides weak enhancements of $T_{C}(p)$ for LaCoAsO and of $T_{N2}(p)$ for SmCoAsO, $T_{N2}$ being the critical temperature of the second successive AFM to AFM transition. A much richer phenomenology is encountered for PrCoAsO, where $p$ is strongly affecting the electrostatic potential of the material and, accordingly, is crucial in determining the statistical occupancy of the crystallographic thermalization sites for muons.

\section{Experimental details}\label{SectChem}

\begin{figure}[t!]
\includegraphics[width=0.47\textwidth]{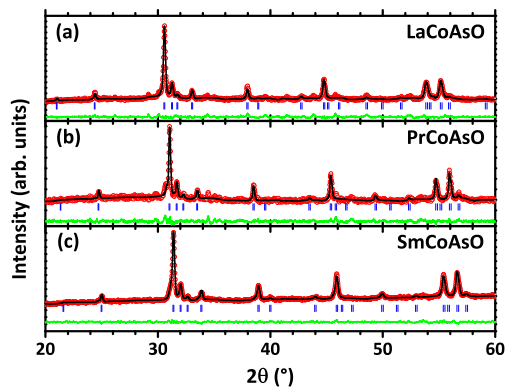}
\caption{(Color online) Observed (red circles) and calculated (blue solid lines) X-ray powder-diffraction patterns at room temperature for the investigated samples of LaCoAsO (see (a) panel), PrCoAsO (see (b) panel) and SmCoAsO (see (c) panel). Black lines are best-fits to experimental data according to a Rietveld analysis.}\label{GraXRD}
\end{figure}
\begin{table}[b!]
\caption{Lattice parameters for the investigated samples after Rietveld refinements of X-ray powder diffraction patterns displayed in figure~\ref{GraXRD}.}
\label{TabLatticeConstants}%
\vspace*{0.3cm}
\bgroup
\begin{center}
\begin{tabular}{cccc}
\hline
\hline
\textbf{Compound} & \textbf{$a$ (\AA)} & \textbf{$c$ (\AA)} & \textbf{Volume (\AA$^{3}$)}\\
\hline
LaCoAsO & $4.048(8)$ & $8.462(7)$ & $138.73(1)$\\
\hline
PrCoAsO & $4.012(6)$ & $8.354(1)$ & $134.51(1)$\\
\hline
SmCoAsO & $3.957(3)$ & $8.242(3)$ & $129.06(1)$\\
\hline
\hline
\end{tabular}
\end{center}
\egroup
\end{table}
Loose powders of $R$CoAsO ($R$ = La, Pr, Sm) were grown via conventional solid-state reactions (see Refs.~\cite{Pal11a,Pal11b} for more details). A Rigaku X-ray diffractometer with Cu K$_{\alpha}$ radiation was employed to investigate and characterize the structural properties of the samples at room temperature ($T$). Diffraction patterns were analyzed by means of the Rietveld method, confirming that all the samples crystallized in the tetragonal phase (space group $P4/nmm$), and the values of the lattice parameters $a$ and $c$ were extracted in turn (see figure~\ref{GraXRD} and table~\ref{TabLatticeConstants} - data for LaCoAsO are taken from \cite{Pra13a}). In agreement with previous reports on $1111$ compounds \cite{Oht09b,Pra13a,Luo10,Nit10}, the full substitution of $R$ progressively shrinks the cell and, accordingly, reduces both $a$ and $c$ parameters with decreasing the ionic radius of $R$ (see table~\ref{TabLatticeConstants}).

Measurements of zero-magnetic-field (ZF) $\mu$SR were performed at the GPD spectrometer ($\mu$E1 beamline) of the S$\mu$S muon source at the Paul Scherrer Institut PSI, Switzerland. Detailed introductions to the $\mu$SR technique(s) can be found in Refs.~\cite{Blu99} and \cite{Yao11}, while \cite{Car13} is a more recent review on $\mu$SR in pnictide materials.  In order to generate nearly-hydrostatic pressures $p \lesssim 2.4$ GPa, a double-wall piston-cylinder cell (PC) made of MP$35$N alloy was employed with Daphne oil 7373 as transmitting medium \cite{Pra13a,Kha11,Dun10}. The PC was loaded at room $T$ and the actual $p$ value was quantified by the shift of the superconducting critical temperature ($T_{c} \sim 3$ K) of a small In manometer inside the cell by means of ac susceptometry. In $\mu$SR experiments under pressure, a high background signal is introduced by the thick walls of the PC, stopping up to $\sim 50$ \% of the incoming muons. For this reason, the asymmetry of the PC must be calibrated in an independent set of experiments as a function of $T$. 

\begin{figure}[b!]
\includegraphics[width=0.48\textwidth]{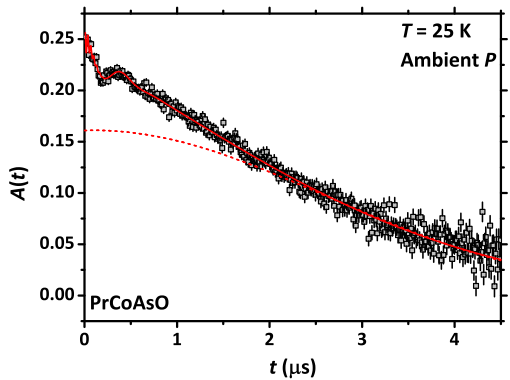}
\caption{(Color online) Repesentative ZF $t$-depolarization for PrCoAsO as measured in the GPD spectrometer (reference measurements in low-background spectrometers are reported in Ref.~\cite{Sug11}). Data are relative to $T = 25$ K and $p = 1$ bar. The red continuous line is a best fit to experimental data according to equation~\ref{EqGeneralFittingZFPCandSample}, while the red dashed line is a guide to the eye showing the expected background contribution from the pressure cell.}\label{GraDepolarization}
\end{figure}

A representative ZF-$\mu$SR asymmetry plot is shown in figure~\ref{GraDepolarization} for the specific case of PrCoAsO. The ZF muon asymmetry $A(t;T)$ reflects the dependence of the spin polarization of the ensemble of implanted muons on time ($t$) and, in turn, the features of the local magnetic interactions leading to its $t$ evolution. Experimental GPD data are fitted for all samples and at all the investigated $T$ values by the general expression
\begin{equation}\label{EqGeneralFittingZFPCandSample}
\frac{A(t;T)}{A_{0}} = f_{PC} \; G_{PC}(t;T) + \left(1 - f_{PC}\right) G_{S}(t;T).
\end{equation}
Here, $A_{0}$ is a calibrated initial asymmetry and the fraction $f_{PC}$ accounts for the incoming muons stopping in the PC, whereas the remaining fraction $\left(1 - f_{PC}\right)$ of muons is implanted directly into the sample and behaves according to the depolarization function $G_{S}(t)$, described in more details in the next section. The empty cell displays fairly constant relaxation parameters within the whole investigated $T$ range, mainly associated with the nuclear magnetic moments of the MP35N alloy. However, in the presence of a FM material this signal is altered by the stray fields from the sample. The details of the more complex PC calibration for FM samples is described elsewhere \cite{Pra13a,Mai11}. Additional ZF-$\mu$SR measurements are performed at the low-background spectrometer Dolly ($\pi$E1 beamline) and are used exclusively as parameter constraints for the fits of equation~\ref{EqGeneralFittingZFPCandSample}. These ZF-$\mu$SR measurements without PC are otherwise similar to those extensively described by Sugiyama et al. in Ref.~\cite{Sug11}.

In the following, in view of the complex PC calibration,  the critical temperatures are measured as \textit{the onset of coherent spin precessions in the transversal component of the sample signal} \cite{Pra13a}. It is noticed that $T_{C}$ values for the current set of samples are systematically lower than what reported previously \cite{Sug11}.

\section{Results and discussion}\label{SectResults}

\subsection{Zero field $\mu$ spin spectroscopy and data analysis}

As anticipated in the previous section, the spin precessions and relaxations in the investigated material are described by the $T$-dependent depolarization function (see equation~\ref{EqGeneralFittingZFPCandSample})
\begin{eqnarray}\label{EqGeneralFittingZFSample}
G_{S}(t;T) & = & G_{p}(t) [1 - v_{m}(T)] + \\ & &  v_{m}(T) \sum_i [f_{i}^{\perp} G_{i}^{\perp}(t;T) + f_{i}^{\parallel} G_{i}^{\parallel}(t)]\nonumber
\end{eqnarray}
where the index $i$ accounts for up to $N = 2$ distinct muon sites in the unit cell. The quantity $v_{m}(T)$ is the magnetic volume fraction of the sample and it accounts for possible inhomogeneous transitions. In the paramagnetic limit $v_{m}(T) = 0$, the muon spin is not relaxed by fast electric fluctuations and only the almost static nuclear dipolar fields contribute to the depolarization $G_{p}(t)$, typically Gaussian with characteristic rate $\sigma_{N}$. Below the critical temperature $T_{C}$, magnetic order yields a non-vanishing local field $\mathbf{B}_i$ at the muon site $i$. In this case, the superscript $\perp$ ($\parallel$) refers to fractions of muons experiencing a local static magnetic field in a transverse (longitudinal) direction with respect to the initial muon spin polarization and $\sum_{i} \left(f_{i}^{\perp} + f_{i}^{\parallel}\right) = 1$. Commensurate long-range, collinear magnetic order yields polarization functions $G_i^\perp(t)$ corresponding to damped precessions at $\omega_{i}=\gamma_{\mu} B_{i} \; (i~=~1,2)$, with Gaussian or exponential relaxation at $T_{2}^{-1}$ rates, reflecting respectively the distribution of local magnetic field values at the muon site or the $\omega_i$ component of their secular spectral density. The longitudinal depolarization fuction is $G_{i}^{\parallel}(t)=e^{-t/T_1} $, probing spin-lattice-like relaxation processes. Equation \ref{EqGeneralFittingZFSample} applies straightforwardly also to the data collected on low-background spectrometers, without PC.
\begin{figure}[b!]
\includegraphics[width=0.47\textwidth]{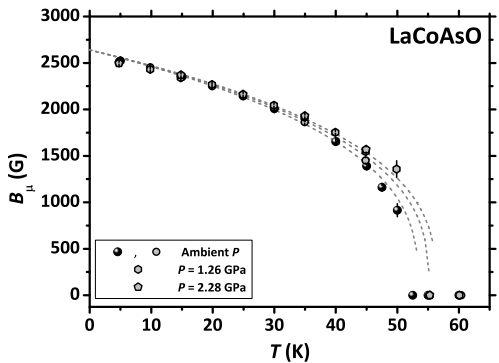}
\caption{$T$ dependence of the internal field at the muon site for LaCoAsO under different conditions of $p$. Filled and empty symbols refer to measurements at Dolly and at GPD with unloaded pressure cell, respectively. The gray dashed lines are best fits to experimental data according to the power-law function described in the text, see equation~\ref{meanfield} (reproduced from \cite{Pra13a}).}\label{GraInternalFieldLaCoAsO}
\end{figure}

For $R$ = La, Pr (and for Sm above $T_{N}$), the samples are ferromagnetic and unmagnetized, but every domain has a net magnetization $M$. The internal field at a muon site is ${\mathbf B}_{i} = {\mathbf B}_{hf} + {\mathbf B}_{d} + 4 \pi {\mathbf M}/3$, where the three terms are respectively the contact hyperfine field, the dipolar field and the Lorentz field \cite{Pra13a,Yao11}. For $R$ = Sm, below $T_{N}$ of course $M = 0$. Since for collinear magnetic structures all three terms are directly proportional to the magnetic moment on Co, the $T$ dependence of the local field may be regarded as a measure of the relative variations of the latter. 

\subsection{LaCoAsO and SmCoAsO}

The results of the application of hydrostatic $p$ up to $2.3$ GPa for LaCoAsO, presented in figure~\ref{GraInternalFieldLaCoAsO}, have been reported previously \cite{Pra13a}, while a detailed description of the ambient $p$ behaviour can be found in Ref.~\cite{Sug11}. Both are summarized here for the aim of clarity. A single oscillating contribution is found ($N = 1$ in equation~\ref{EqGeneralFittingZFSample}) at all the $T$ and $p$ values where measurements are performed. The $T$ dependence of the internal field at the muon site $B(T)$ can be well described by a power-law trend
\begin{equation}\label{meanfield}
B(T)=B_{0}\left(1-\frac{T}{T_{C}}\right)^\beta
\end{equation}
with $\beta = 0.34$ characteristic of the FM ordering of the Co sublattice \cite{Pra13a}, namely the only magnetic species inside the material, and possibly denoting $3$D magnetic correlations \cite{Kau85}. By keeping $\beta = 0.34$ as fixed parameter, the $T_{C}$ value can be extracted as a fitting parameter from data shown in figure~\ref{GraInternalFieldLaCoAsO}. Results clearly display $T_{C} = 53.5$ K at ambient $p$, a value that is slightly enhanced to $56.5$ K for $p \simeq 2.3$ GPa (see figure~\ref{BTp} later on). When compared to the much stronger response in the isostructural $R$CoPO compounds, this enhancement is clearly negligible \cite{Pra13a}. It should be stressed that, at variance with the isostructural compound LaCoPO, no dependence of $B_{0}$ on $p$ is reported in the current case (see figure~\ref{BTp} later on).

\begin{figure}[b!]
\includegraphics[width=0.47\textwidth]{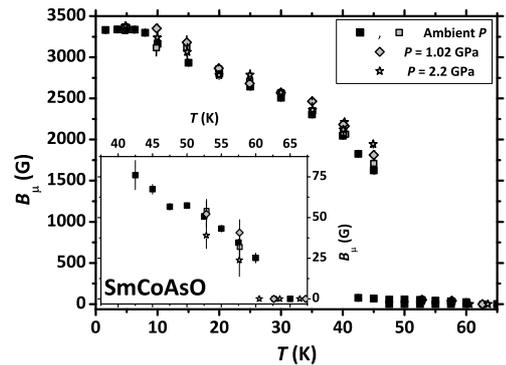}
\caption{$T$ dependence of the internal field(s) at the muon site(s) for SmCoAsO under different conditions of $p$. Filled and empty symbols refer to measurements at Dolly and at GPD with unloaded pressure cell, respectively. Data in the intermediate $T$ region are enlarged in the inset for the aim of clarity. From these data and within the experimental error, no effect of $p$ can be detected at all.}\label{GraInternalFieldSmCoAsO}
\end{figure}
Data relative to the internal field(s) in SmCoAsO are reported in figure~\ref{GraInternalFieldSmCoAsO}.  A discontinuity is observed in the region $42$ K $\lesssim T \lesssim 46$ K, the value of the internal magnetic field above and below that region being different by one order of magnitude. This was already interpreted in Ref.~\cite{Sug11} as a progressive FM to AFM transition with lowering $T$, as also supported by dc magnetization. The AFM nature of the high internal field, low-$T$ phase is also witnessed by the sudden suppression of the high relaxation of the PC signal across the transition, due to the disappearance of the stray fields from the FM sample (data not shown). For $42$ K $\lesssim T \lesssim 46$ K  macroscopic coexistence of AFM and FM phases is indicated by the presence of two characteristic precessing components. This confirms that the muon occupies a unique site across the whole $T$ range. 

Similarly to the case of LaCoAsO, $p$ has a very modest effect both on $T_{C}$ and on $T_{N1} \sim 45$ K, the variation of the latter being smaller than the width of the coexistence region. Within the experimental error there is no variation of both internal fields, therefore no detectable pressure effect on the magnetic structures and on the value of the Co and Sm magnetic moments. Concerning the low-$T$ AFM to AFM transition (likely the onset of Sm order) at $T_{N2} \sim 15$ K, our results actually show a slight pressure variation in that region, indicating that $p$ may be increasing the Sm coupling to the Co lattice.

\subsection{PrCoAsO: muon diffusion}

\begin{figure}[t!]
\includegraphics[width=0.44\textwidth]{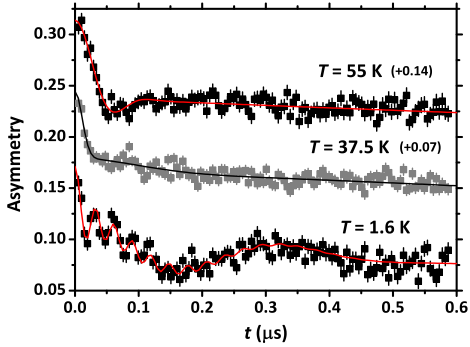}
\caption{(Color online) Early-times ZF $\mu$SR asymmetry in PrCoAsO at three representative temperatures and ambient pressure on Dolly. The solid lines are best fits to equation~\ref{EqGeneralFittingZFSample} (see text)}\label{AmbientPressurePrCoAsOAsym}
\end{figure}
\begin{figure}[t!]
\includegraphics[width=0.44\textwidth]{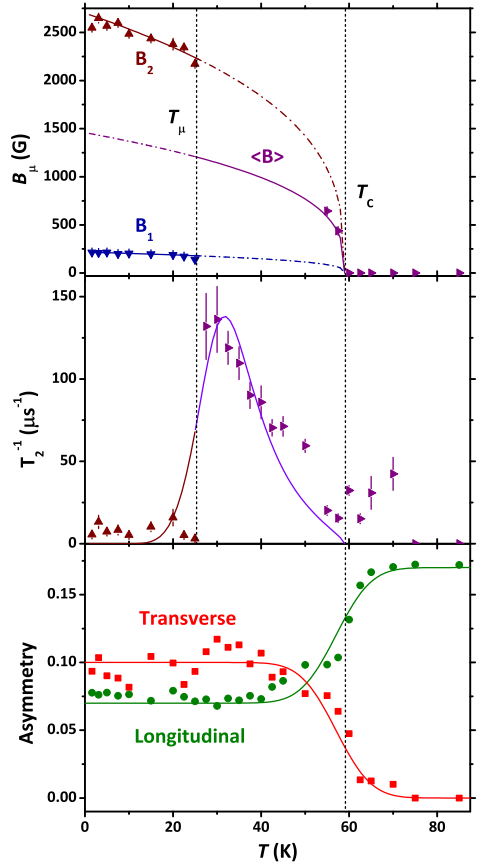}
\caption{(Color online) Temperature dependence of ZF $\mu$SR parameters in PrCoAsO at ambient pressure on Dolly. Bottom panel: Transverse and longitudinal components of the muon asymmetry; the paramagnetic-ferromagnetic transition at $T_C$ is marked by a vertical dash-dotted line. Central panel: transverse relaxation rate $T_2^{-1}$, $\blacktriangle$, $T<T_\mu$, for the high field site $B_2$ and $\blacktriangleright$, $T>T_\mu$, for the average internal field; the crossover to a diffusing muon regime is indicated by a vertical dashed line at $T_\mu$; the curve is the relaxation rate expected for a diffusing muon (see text). Top: internal fields at the muon sites, $\blacktriangleleft$, $T<T_\mu$, for the low field site $B_1$ besides the two already mentioned above; The curve is a fit to a mean field dependence (see text).}\label{AmbientPressurePrCoAsO}
\end{figure}
The case of PrCoAsO requires a detailed description of the ambient $p$ data, which is better done on those obtained at the low-background spectrometer Dolly. Our data reproduce the main features already described in \cite{Sug11}. The $t$ dependence of the asymmetry reported in figure~\ref{AmbientPressurePrCoAsOAsym} and the best fit parameters summarized in figure~\ref{AmbientPressurePrCoAsO} illustrate two distinct behaviors. At $T = 1.6$ K and in general for $T \lesssim T_{\mu} = 25$ K two distinct internal fields $B_{1}(T)$ and $B_{2}(T)$ can be distinguished, whose saturation values are different by at least one order of magnitude. By analogy with RFeAsO \cite{Mae09,DeR12} and $R$CoPO \cite{Pra13a}, where a detailed DFT investigation was performed, these two sites are identified with the two minima of the Coulomb potential close to the FeAs (CoP) layer and close to the $R$O layer, respectively. In contrast, for $T_\mu\lesssim T \lesssim 60$ K, ($T = 37.5$ K and $55$ K in figure~\ref{AmbientPressurePrCoAsOAsym}) a single transverse component of the asymmetry shows an overdamped oscillation at a unique field which can be resolved only very near to $60$ K. The lower panel of the figure shows the jump of the longitudinal and transverse asymmetries at $T_{C}$, indicating the appearance of an internal field at the muon site below that temperature. The solid lines, followed by the data within minor departures, show the asymmetry ratio expected for a magnetically ordered polycrystalline sample (with minor background contributions).

The detailed origin of the discontinuity at $T_{\mu}$ is not discussed in \cite{Sug11}. However, it should be remarked that no anomaly can be discerned in the dc magnetization data \cite{Sug11}, suggesting that such feature should be associated to the physics of the positively charged $\mu$ in the crystalline environment of PrCoAsO, rather than to a change in the magnetic properties of the material. Notably, the internal field $B(T)$ for $T \simeq 55 - 60$ K is very close to the expected average of $B_{1}(T)$ and $B_{2}(T)$, namely $\langle B\rangle = [B_{2}(T) + B_{1}(T)]/2$. This can be demonstrated by comparing their temperature dependence for $T \lesssim T_{\mu}$ with power-law trends (see equation~\ref{meanfield}), with $(B_{1,2})_{0} \simeq 220,\, 2700$ G and $\beta = 0.34$, represented by the solid curves in figure~\ref{AmbientPressurePrCoAsO}, top panel. Their extensions for $T \gtrsim T_{\mu}$ are denoted by a dash-dotted line, while the curve that becomes solid above $T_{\mu}$ is the average $\langle B(T)\rangle$, that we identify from now on with the single internal field detected for $T\gtrsim 55 $ K. It must be noted that a finite $\langle B(T)\rangle$ value is expected to exist because of the FM magnetic structure throughout the entire ordered phase, as indicated by dc magnetization \cite{Sug11}. In contrast diffusion in an AF structure where local fields alternate at opposite values would result in a vanishing average field.

This scenario agrees with $T_{\mu}$ being the onset of muon diffusion by thermally activated jumps between sites $1$ and $2$. Assuming an activated residence time $\tau(T)= \tau_\infty \times \exp(T_{A}/T)$ for this dynamical process and the time-honoured Bloembergen, Purcell, Pound \cite{Blo48} mechanism, the relaxation rate develops a peak
\begin{equation}\label{BPP}
\frac{1}{T_{2}(T)} = \frac{[\omega_{1}(T)-\omega_{2}(T)]^{2} \tau(T)}{1+\langle\omega(T)\rangle^{2}\tau^{2}(T)},
\end{equation} 
where $\omega_{i}(T) = 2 \pi \gamma_\mu B_{i}(T)$ and $\langle\omega(T)\rangle = 2 \pi \gamma_\mu \langle B(T)\rangle$. This peak is shown by the solid curve in the central panel of figure~\ref{AmbientPressurePrCoAsO} for best fit values $\tau_\infty = 1 \times 10^{-10}$ s and $T_{A} = 150$ K. These are reasonable values for a thermally-activated diffusion process. This indicates that indeed muon diffusion sets in with these activation parameters among sites characterized by the same direction of the local field in the FM structure of PrCoAsO.

\begin{figure}[t!]
\includegraphics[width=0.23\textwidth]{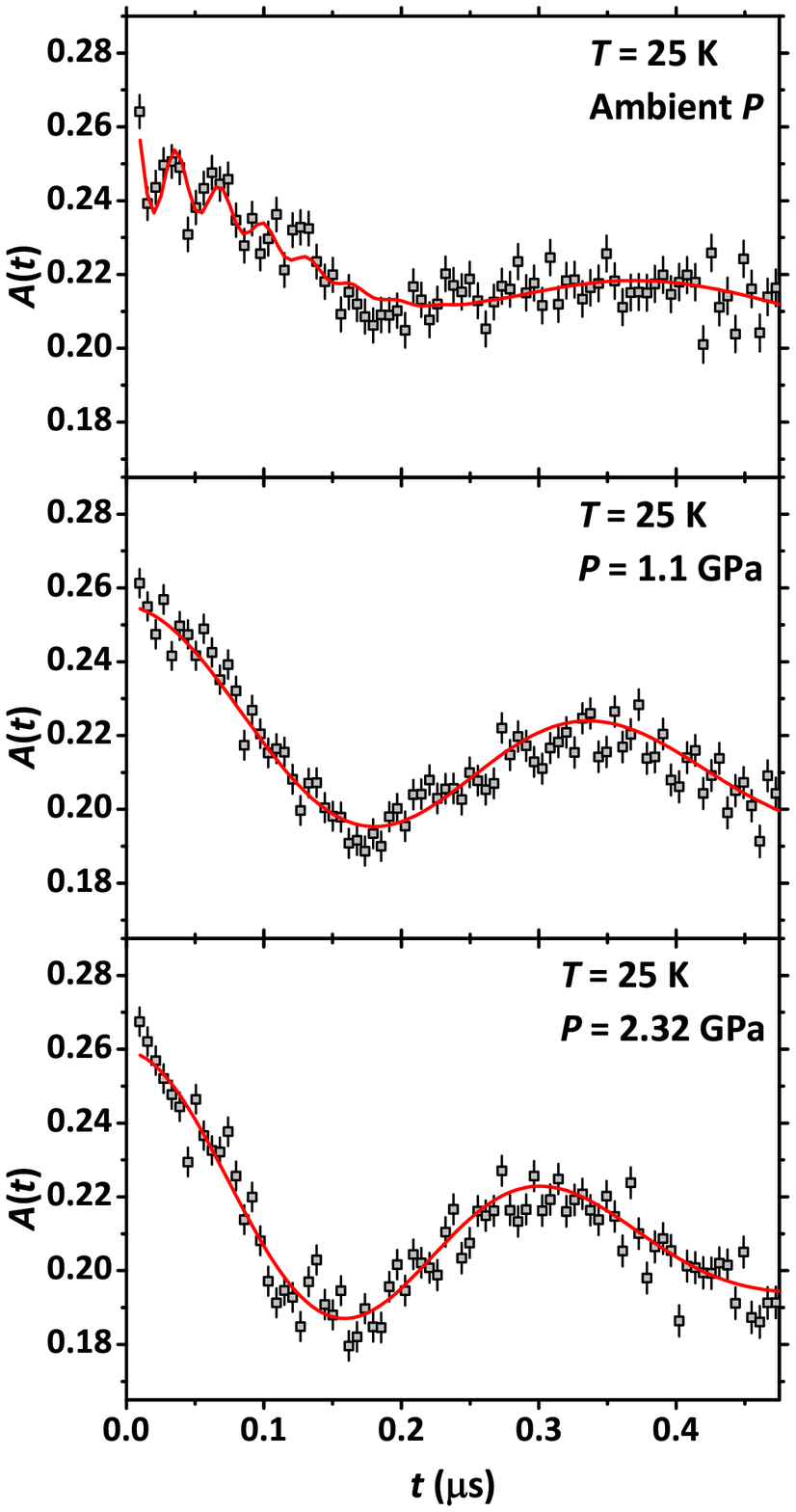} \hfill \includegraphics[width=0.23\textwidth]{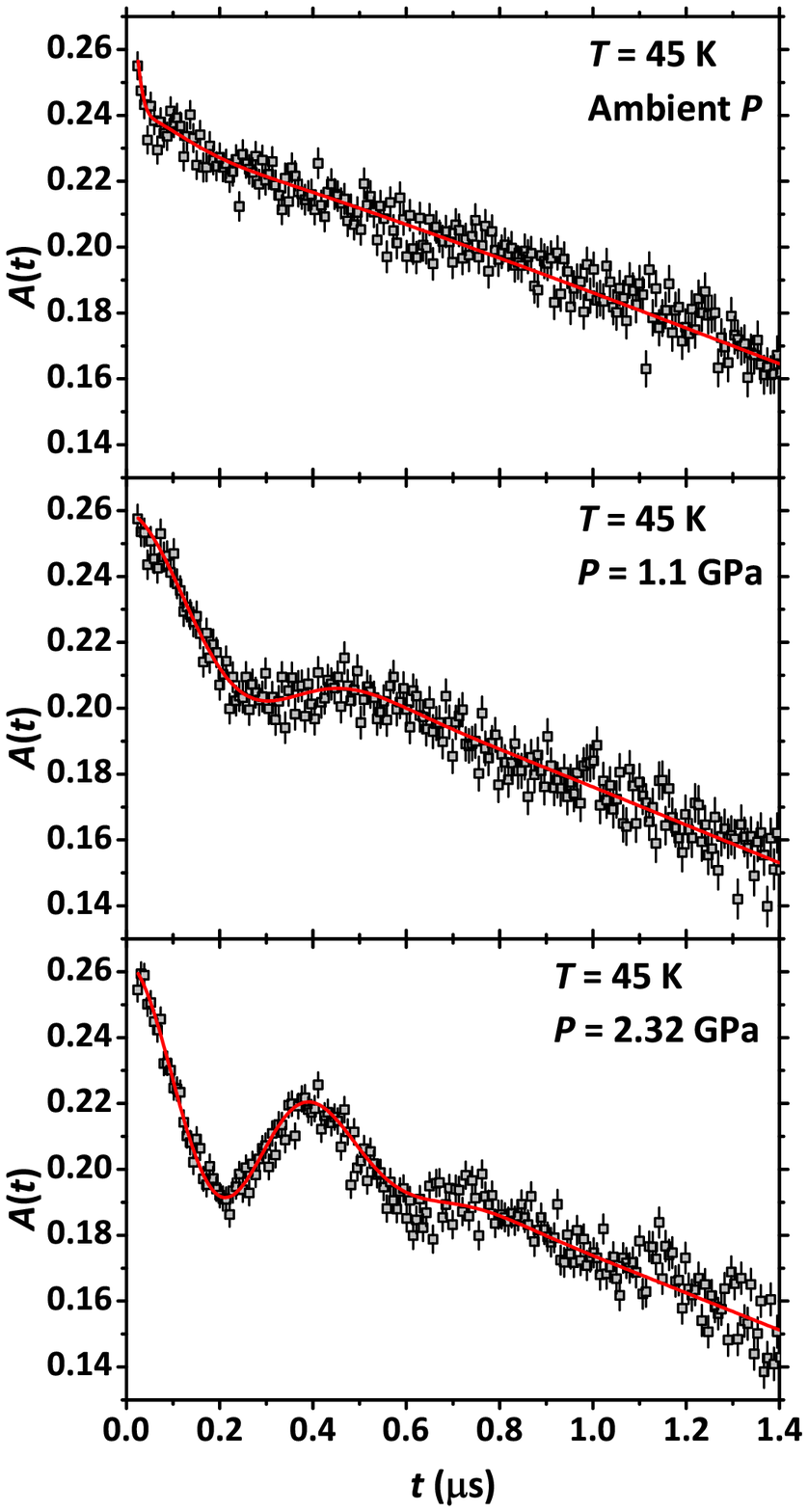}
\caption{(Color online) Early-time asymmetries for PrCoPO. Data on the left (right) column are relative to $T = 25$ K ($T = 45$ K, notice the different time scale), while $p$ is increasing in the vertical direction).}\label{GraDepolarizationVsP}
\end{figure}
\begin{figure}[b!]
\includegraphics[width=0.47\textwidth]{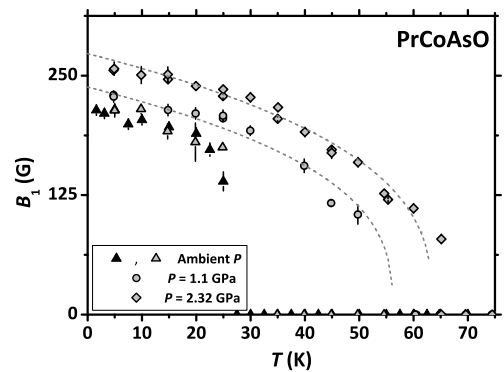} 
\caption{(Color online) $T$ dependence of the internal field at the muon site ``$1$'' for PrCoAsO at $p = 0, 1.1$ and $2.3$ GPa. Filled and empty symbols refer to measurements at Dolly and at GPD with unloaded pressure cell, respectively. Ambient pressure data without cell (Dolly) are shown for comparison. The dashed lines are best fits to experimental data according to equation~\ref{meanfield}.}\label{GraInternalFieldPrCoAsO}
\end{figure}
Let us now focus on the dependence of the internal field on $p$. At first glance it is apparent in figure~\ref{GraDepolarizationVsP} that at $T=25$ K $< T_\mu$ pressure suppresses the high frequency site 2 and increases the fraction of the slow frequency site 1, whereas at $T=45$ K  $>T_\mu$ pressure suppresses the initial fast decaying signal and restores the low frequency site 1. Since the fast decay at the average internal field $\langle B\rangle$ is due to muon diffusion, this is an indication that pressure stabilizes site 1. In addition, the low frequency increases on increasing pressure in each vertical series at constant temperature, and this indicates that pressure enhances magnetic order.

Summarizing, the main effect of $p$ is that of quenching the diffusion process described above by stabilizing muon site 1. Site 2 becomes energetically unfavourable by the compression of the lattice and it is completely depopulated already at $p = 1.1$ GPa. Accordingly, we concentrate on the $T$ dependence of $B_{1}(T)$ for the different conditions of $p$ ($0, 1.1, 2.3$ GPa), shown in figure~\ref{GraInternalFieldPrCoAsO}. We stress that, whereas for $p = 0$ the full field dependence is that of figure \ref{AmbientPressurePrCoAsO}, for $p\ge 1.1$ GPa the unique muon local field is $B_{1}$. The same power-law curve described by equation~\ref{meanfield} provides an estimate of the $T = 0$ local field values at muon site 1, $(B_{1})_{0}(p)$ and of the transition temperatures $T_{C}(p)$, assuming the standard value of $\beta = 0.34$.

\section{Discussion and conclusions}
\begin{figure}[t!]
\includegraphics[width=0.23\textwidth]{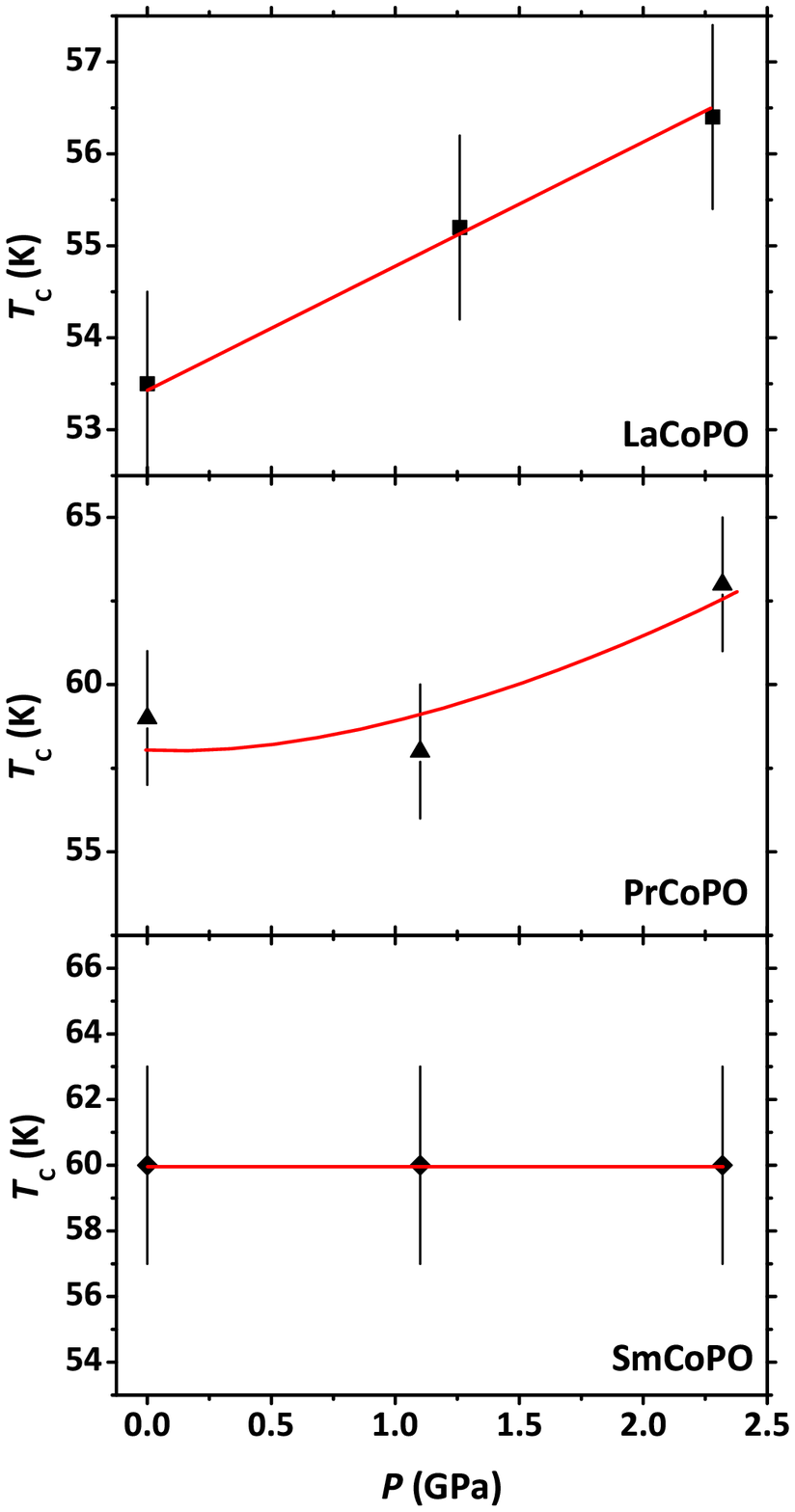} \hfill \includegraphics[width=0.24\textwidth]{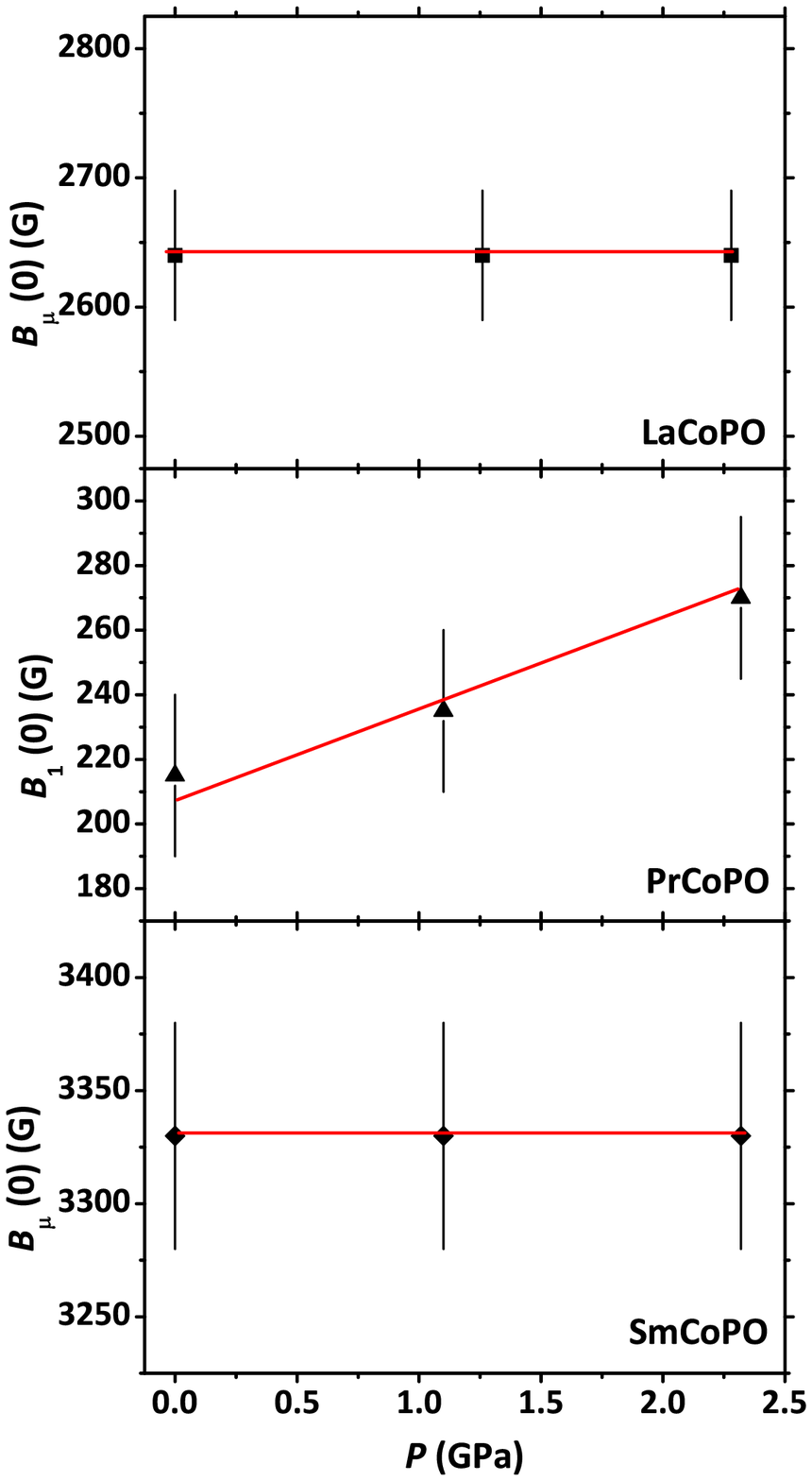}
\caption{(Color online) Summarizing dependence on $p$ of the critical transitions temperatures $T_{C}$ to the FM phase (left column) and of the saturation values for the internal magnetic field at the muon site (right column).}\label{BTp}
\end{figure}

A summarizing pressure dependence of the zero temperature local field, $B_{\mu0}(p)$, and of the ordering temperatures $T_C(p), T_N(p)$ is shown in figure~\ref{BTp}. LaCoAsO and SmCoAsO have been shown to be fairly ``magnetically hard'' materials, meaning that pressure is not playing an important role in influencing their physical properties. In LaCoAsO, only a weak dependence of $T_{C}$ on $p$ is detected \cite{Pra13a} (see figure~\ref{BTp}). On the other side, in SmCoAsO two different AFM phases are suggested in Ref.~\cite{Sug11}, reporting a modification of the magnetic structure at $T_{N2} \sim 15$ K, likely the onset of Sm order. Our results actually show a slight pressure variation in that region, indicating that $p$ may be increasing the Sm coupling to the Co lattice (see figure~\ref{GraInternalFieldSmCoAsO}).

In the case of PrCoAsO, the agreement of our data with power-law curves with $\beta = 0.34$ (see equation~\ref{meanfield}) shows that Pr does not order down to $T = 1.6$K. This confirms and extends the indications of Ref.~\cite{Tiw14}. Our ambient pressure $\mu$SR results alter former conclusions \cite{Sug11}, where two discontinuities were identified in the temperature dependence of the muon signal and two further magnetic transitions within a FM-type structure were considered as a possible cause. As we have shown in figure~\ref{AmbientPressurePrCoAsO} there is no additional magnetic transition, in agreement with the smooth dc magnetic susceptibility \cite{Sug11,Tiw14}. Instead, one discontinuity at $T_{\mu} = 25$ K is due to the onset of muon diffusion, whereas the second one at higher temperature is just the reappearance of a discernible average precession frequency when the jump rate becomes sufficiently fast. The observation of the simple average of the two low temperature fields indicates that the sites have equal residence times in the diffusion. However, the high field muon site is unstable under pressure and its population vanishes already at $1.1$ GPa. In order to determine the effect of pressure on magnetic properties by $\mu$SR, one must therefore concentrate on site 1. In contrast, $R$ = La and Sm always show a single muon site and no change under pressure.

In summary, we performed detailed measurements by means of muon spin spectroscopy on $R$CoAsO, with $R$ = La, Pr and Sm, under applied hydrostatic
pressure. Our results demonstrate that in these materials it is important to check that the muon site is not influenced by pressure and temperature. La and Sm compounds show small variations with pressure, whereas Pr shows a much larger response, possibly connected with the fact that Pr is accidentally magnetically inactive at ambient pressure.

\section{Acknowledgements}

GP acknowledges support from the Alexander von Humboldt Post-Doc Fellowship Program. PC  and SS acknowledge support from Fondazione Cariplo (research grant No. 2011-0266). RDR, PB, PC and SS ackowledge partial support of PRIN2012 project 2012X3YFZ2.



\end{document}